\documentclass[onecolumn,prb]{revtex4}
\usepackage{graphicx}
\usepackage{dcolumn}
\usepackage{bm}
\usepackage{amsmath}

\begin{document}

\preprint{}
\title{Theory of Josephson effect in unconventional superconducting
junctions with diffusive barriers}
\author{T. Yokoyama$^1$, Y. Tanaka$^1$ and A. A. Golubov$^2$}
\affiliation{$^1$Department of Applied Physics, Nagoya University, Nagoya, 464-8603, Japan%
\\
and CREST, Japan Science and Technology Corporation (JST) Nagoya, 464-8603,
Japan \\
$^2$ Faculty of Science and Technology, University of Twente, 7500 AE,
Enschede, The Netherlands}
\date{\today}

\begin{abstract}
We study theoretically the Josephson effect in junctions based on
unconventional superconductors with diffusive barriers, using the
quasiclassical Green's function formalism. Generalized boundary
conditions at junction interfaces applicable to unconventional
superconductors are derived by calculating a matrix current within
the circuit transport theory. Applying these boundary conditions, we
have calculated the Josephson current in structures with various
pairing symmetries. A number of predictions are made: (a)
nonmonotonic temperature dependence in $d$-wave
superconductor/diffusive normal metal/$d$-wave superconductor
(D/DN/D) junctions, (b) anomalous
current-phase relation in $p $-wave superconductor/diffusive normal metal/$p$%
-wave superconductor (P/DN/P) junctions, (c) second harmonics in D/DN/D and
P/DN/P junctions, (d) a double peak structure of the critical current in
D/DF/D junctions, (e) enhanced Josephson current by the exchange field in
S/DF/P junctions. We have also investigated peculiarities of the Josephson
coupling in D/DF/D, P/DF/P and S/DF/P junctions. An oscillatory behavior of
the supercurrent and the second harmonics in the current-phase relation are
studied as a function of the length of the diffusive ferromagnet.
\end{abstract}

\maketitle

%$^3$Department of Applied Physics, Hokkaido University,Sapporo, 060-8628,
%Japan}

% and  J. Inoue$^1$

%--- title ---

%--- author ---

%
%--- address ---

%
%--- date ---

%   Abstract
%-----------------------------------------------------------

%-----------------------------------------------------------


\section{Introduction}

%-----------------------------------------------------------
The Josephson effect\cite{Josephson} has been studied in various types of
junctions\cite{deGennes,Likharev,Golubov} motivated by fundamental interest
and potential applications for future technology. In superconductor /
diffusive normal metal / superconductor (S/DN/S) junctions the critical
current increases monotonically with decreasing temperature\cite%
{Likharev,Kupriyanov,Zaitsev,Golubov} because proximity effect is enhanced
at low temperatures. Superconducting junctions with ferromagnetic
interlayers have shown rich physics due to the interplay of proximity effect
and the exchange field\cite{buzdinrev,bverev}. When DN is replaced by a
diffusive ferromagnet (DF), it was predicted that $\pi $ junctions can be
realized\cite{Bulaevskii,Buzdin,Buzdin2,Demler,Fominov,Buzdin3,Houzet,Faure}%
. The physical reason for a $\pi $-state is nonzero momentum of induced
Cooper pairs in the ferromagnet \cite{Demler}, similar to the so-called
Fulde-Ferrel-Larkin-Ovchinnikov state \cite{Fulde,Larkin} in a magnetic
superconductor. SFS $\pi $ junctions were realized experimentally by several
groups\cite%
{Ryazanov,Kontos,Sellier,Frolov,Blu04,Su02,Be02,She06,Wei06,Pepe,Born}.

% An application of $\pi$-junctions to quantum bit was proposed\cite{Ioffe}.
%For applications, it is better to use high temperature ($d$-wave) superconducto%rs.

In $d$-wave superconductor junctions, one of the most remarkable phenomena
is the formation of midgap Andreev resonant states (MARS) at interfaces\cite%
{Buch}. The MARS stem from sign change of pair potentials of $d$-wave
superconductors \cite{TK95}. In $d$-wave superconductor / insulator / $d$%
-wave superconductor junctions, $\pi $-junctions emerge due to the formation
of the MARS\cite{Barash,Golubov2}. In order to clarify the role of proximity
effect and MARS, Tanaka \textit{et al.} have extended the circuit theory%
\cite{Nazarov2} to the junctions with unconventional superconductors \cite%
{Nazarov2003,TNGK,p-wave}. The conservation of matrix current enables one to
apply the generalized Kirchhoff's rules to unconventional superconducting
junctions and to derive the boundary conditions for the Usadel equation\cite%
{Usadel} widely used in diffusive superconducting junctions. Application of
this theory to the DN/$d$-wave superconductor (DN/D) junctions has shown
that the formation of MARS strongly competes with the proximity effect in DN%
\cite{Nazarov2003,TNGK}. It was also demonstrated that the formation of MARS
coexists with the proximity effect in DN/$p$-wave superconductor (DN/P)
junctions, which produces a giant zero bias conductance peak (ZBCP)\cite%
{p-wave}.

Recently this theory has been extended to diffusive Josephson junctions with
unconventional superconductors\cite{Yoko}. It is clarified that a
nonmonotonic temperature dependence of the critical current appears in
D/DN/D junctions due to the competition between the proximity effect and the
formation of MARS. Concerning unconventional superconducting junctions with
clean ferromagnet, Josephson effect is studied in Ref.\cite{TK}. However
Josephson effect in unconventional superconducting junctions with diffusive
ferromagnet has not yet been studied, although proximity effect in these
junctions was studied recently\cite{Yokoyama}. Moreover detailed derivation
of the boundary conditions is not given and only D/DN/D junctions are
considered in Ref.\cite{Yoko}. Since proximity effect and MARS strongly
influence the density of states\cite{Nazarov2003,TNGK,p-wave}, they should
also crucially influence Josephson effect in diffusive unconventional
superconducting junctions.

The purpose of the present paper is to study the Josephson effect in various
types of conventional and unconventional superconducting junctions with DN
or DF interlayers. We first provide the technical details of derivation in
Ref.\cite{Yoko}. Then, solving the Usadel equations, we apply the general
approach to study the influence of the exchange field, the proximity effect
and the formation of MARS on Josephson current simultaneously. A number of
peculiarities in the Josephson current are found depending on the pairing
symmetry: a nonmonotonic temperature dependence in D/DN/D junctions,
anomalous current-phase relation in P/DN/P junctions, second harmonics in
the current-phase relation and their half-periodic oscillations as a function of the length of DF in D/DN/D and P/DN/P junctions, transitions to a $%
\pi$ state in D/DF/D, P/DF/P and S/DF/P junctions, a double peak structure
in the temperature dependence of the critical current in D/DF/D junctions,
and enhancement of Josephson current by the exchange field in S/DF/P
junctions.

\section{Formulation}

We consider a junction consisting of unconventional superconductors
 (USCs) connected by a quasi-one-dimensional DN (or DF) with a
resistance $R_{d}$ and a length $L$ much larger than the mean free path. The
DN/USC interface located at $x=0$ has the resistance $R_{b}^{\prime }$, while
the DN/USC interface located at $x=L$ has the resistance $R_{b}$. We model
infinitely narrow insulating barriers by the delta function $U(x)=H\delta
(x-L)+H^{\prime }\delta (x)$. %\begin{figure}[htb]
%\begin{center}
%\scalebox{0.4}{
%\includegraphics[width=22.0cm,clip]{fig0.eps}}
%\end{center}
%\caption{ Schematic illustration of the model.}
%\label{f0}
%\end{figure}
The resulting transparencies of the junctions $T_{m}$ and $T_{m}^{\prime }$
are given by $T_{m}=4\cos ^{2}\phi /(4\cos ^{2}\phi +Z^{2})$ and $%
T_{m}^{\prime }=4\cos ^{2}\phi /(4\cos ^{2}\phi +{Z^{\prime }}^{2})$, where $%
Z=2H/v_{F}$ and $Z^{\prime }=2H^{\prime }/v_{F}$ are dimensionless
constants, $\phi $ is the injection angle measured from the interface normal
to the junction and $v_{F}$ is Fermi velocity.
%%%%%%%%%%%%%%%%%%%%%%%%%%%%%%%%%%%%%%%%%%%%%%%%%%%%%%%%%%%%
%%%%%%%%%%%%%%%%%%%%%%%%%%%%%%%%%%%%%%%%%%%%%%%%%%%%%%%%%%%%%

%%%%%%%%%%%%%%%%%%%%%%%%%%%%%%%%%%%%%%%%%%%%%%
% Usadel equation in DF
%%%%%%%%%%%%%%%%%%%%%%%%%%%%%%%%%%%%%%%%%%%%%%%

In order to study the Josephson effect in diffusive USC junctions, we first
concentrate on the quasiclassical Keldysh-Nambu Green's function in DN
defined by $\check{G}_{N}(x)$. Its retarded part $\hat{R}_{N}(x)$ can be
expressed as%
\begin{equation}
\hat{R}_{N}(x)=\cos \psi \sin \theta \hat{\tau}_{1}+\sin \psi \sin \theta
\hat{\tau}_{2}+\cos \theta \hat{\tau}_{3},
\end{equation}%
with Pauli matrices in the electron-hole space, $\hat{\tau}_{1}$, $\hat{\tau}%
_{2}$, and $\hat{\tau}_{3}$. Since $\hat{R}_{N}(x)$ obeys the Usadel
equation, following equations are satisfied,
\begin{equation}
D[\frac{\partial ^{2}}{\partial x^{2}}\theta -(\frac{\partial \psi }{%
\partial x})^{2}\cos \theta \sin \theta ]+2i(\varepsilon +(-)h)\sin \theta
=0,
\end{equation}%
\begin{equation}
\frac{\partial }{\partial x}[\sin ^{2}\theta (\frac{\partial \psi }{\partial
x})]=0
\end{equation}%
for majority (minority) spin with the diffusion constant $D$ and exchange
field $h$. The boundary condition of $\check{G}_{N}(x)$ at DN/USC interface
is given by\cite{Yoko,p-wave}
\begin{equation}
\frac{L}{R_{d}}[\check{G}_{N}(x)\frac{\partial \check{G}_{N}(x)}{\partial x}%
]_{\mid x=L_{-}}=-\frac{<\check{I}(\phi )>}{R_{b}},
\end{equation}

\begin{equation}
\check{I}(\phi)=2[\check{G}_{1},\check{B}],
\end{equation}

\begin{equation}
\check{B}=(-T_{1}[\check{G}_{1},\check{H}_{-}^{-1}]+\check{H}_{-}^{-1}\check{%
H}_{+}-T_{1}^{2}\check{G}_{1}\check{H}_{-}^{-1}\check{H}_{+}\check{G}%
_{1})^{-1}(T_{1}(1-\check{H}_{-}^{-1})+T_{1}^{2}\check{G}_{1}\check{H}%
_{-}^{-1}\check{H}_{+})
\end{equation}%
with $\check{G}_{1}=\check{G}_{N}(x=L_{-})$, $\check{H}_{\pm }=(\check{G}%
_{2+}\pm \check{G}_{2-})/2$, and $T_{1}=T/(2-T+2\sqrt{1-T})$, where $\check{G%
}_{2\pm }$ is the asymptotic Green's function in USC as defined in our
previous papers\cite{p-wave}. The average over the various angles of
injected particles at the interface is defined as
\begin{equation}
<\check{I}(\phi )>=\int_{-\pi /2}^{\pi /2}d\phi \cos \phi \check{I}(\phi
)/\int_{-\pi /2}^{\pi /2}d\phi T(\phi )\cos \phi  \label{average}
\end{equation}%
with $\check{I}(\phi )=\check{I}$ and $T(\phi )=T$. The resistance of the
interface $R_{b}$ is given by
\begin{equation}
R_{b}=\frac{2R_{0}}{\int_{-\pi /2}^{\pi /2}d\phi T(\phi )\cos \phi }
\end{equation}%
with Sharvin resistance at the interface, $R_{0}$. Retarded components of $%
\check{G}_{1}$ and $\check{G}_{2\pm }$ are given by $\hat{R}_{1,2\pm }$
where $\hat{R}_{2\pm }$ is expressed by $\hat{R}_{2\pm }=g_{\pm }\hat{\tau}%
_{3}+f_{\pm }\hat{\tau}_{2}$ with $g_{\pm }=\varepsilon /\sqrt{\varepsilon
^{2}-\Delta _{\pm }^{2}}$ and $f_{\pm }=\Delta _{\pm }/\sqrt{\Delta _{\pm
}^{2}-\varepsilon ^{2}}$, where $\varepsilon $ denotes the quasiparticle
energy measured from the Fermi energy. $\Delta _{+}$ $(\Delta _{-})$ is the
effective pair potential felt by quasiparticles with an injection angle $%
\phi $ $(\pi -\phi )$. We also denote $\hat{R}_{p},\hat{R}_{m}$ and $\hat{I}%
_{R}$ as retarded part of $\check{H}_{+}$, $\check{H}_{-}$ and $\check{I}$.

Now we discuss the boundary condition of the retarded part of Keldysh-Nambu
Green's function at DN/USC interface. The left side of the boundary
condition of Eq. (4) can be expressed as
\begin{equation}
\frac{L}{R_{d}}\hat{R}_{N}(x)\frac{\partial }{\partial x}\hat{R}%
_{N}(x)|_{x=L}=\frac{Li}{R_{d}}[\left( {\ -\frac{{\partial \theta }}{{\theta
x}}\sin \psi -\frac{{\partial \psi }}{{\theta x}}\sin \theta \cos \theta
\cos \psi }\right) \hat{\tau}_{1}+\left( {\frac{{\partial \theta }}{{\theta x%
}}\cos \psi -\frac{{\partial \psi }}{{\theta x}}\sin \theta \cos \theta \sin
\psi }\right) \hat{\tau}_{2}+\frac{{\partial \psi }}{{\theta x}}\sin
^{2}\theta \hat{\tau}_{3}].
\end{equation}

In the right side of Eq. (4), $\hat{I}_{R}$ can be expressed by using
several spectral vectors:
\begin{equation*}
\hat{I}_{R}=4iT_{1}(\bm{ d}_{R} \cdot \bm{ d}_{R})^{-1} \{ -\frac{1}{2}(1 +
T_{1}^{2})({\bm s}_{2+}-{\bm s}_{2-})^{2} [{\bm s}_{1}\times({\bm s}_{2+} + {%
\bm s}_{2-})]\cdot \hat{\bm{\tau}}  \label{spectral}
\end{equation*}
\begin{equation*}
+2T_{1}{\bm s}_{1}\cdot({\bm s}_{2+} \times {\bm s}_{2-}) [{\bm s}_{1}
\times ({\bm s}_{2+} \times {\bm s}_{2-})] \cdot \hat{\bm{\tau}}
\end{equation*}
\begin{equation*}
+2T_{1}{\bm s}_{1}\cdot({\bm s}_{2+} - {\bm s}_{2-}) [{\bm s}_{1} \times ({%
\bm s}_{2+} - {\bm s}_{2-})] \cdot \hat{\bm{\tau}}
\end{equation*}
\begin{equation*}
-i(1+T_{1}^{2})(1 -{\bm s}_{2+}\cdot{\bm s}_{2-}) [{\bm s}_{1} \times ({\bm s%
}_{2+} \times {\bm s}_{2-})]\cdot \hat{\bm{\tau}}
\end{equation*}
\begin{equation}
+2iT_{1}(1 -{\bm s}_{2+}\cdot{\bm s}_{2-}) [{\bm s}_{1}\cdot({\bm s}_{2+}-{%
\bm s}_{2-}){\bm s}_{1} -({\bm s}_{2+} -{\bm s}_{2-})]\cdot \hat{\bm{\tau}}
\},
\end{equation}
\begin{equation}
\bm{d}_{R}=(1 + T_{1}^{2})(\bm{s}_{2+} \times \bm{s}_{2-}) -2T_{1}\bm{s}%
_{1}\times(\bm{s}_{2+}-\bm{s}_{2-}) -2T_{1}^{2}\bm{s}_{1}\cdot(\bm{s}%
_{2+}\times\bm{s}_{2-})\bm{s}_{1}
\end{equation}
with $\hat{R}_{1}={\bm s}_{1}\cdot\hat{\bm{\tau}}$ and $\hat{R}_{2\pm}={\bm s%
}_{2\pm}\cdot\hat{\bm{\tau}}$\cite{p-wave}.

The spectral vectors ${\bm s}_{1}$ and ${\bm s}_{2\pm }$ are given by
\begin{equation}
s_{1}=\left( {\begin{array}{*{20}c} {\sin \theta \cos \psi } \\ {\sin \theta
\sin \psi } \\ {\cos \theta } \\ \end{array}}\right) ,s_{2\pm }=\left( {%
\begin{array}{*{20}c} {f_ \pm \cos \Psi } \\ { f_ \pm \sin \Psi } \\ {g_ \pm
} \\ \end{array}}\right)
\end{equation}%
where $\Psi $ denotes the phase of the USC. Then $\hat{I}_{R}$ is reduced to
\begin{equation}
\begin{array}{l}
\hat{I}_{R}=2iT\left[ {\left( {2-T}\right) +T\left\{ {\bar{f}_{S}\sin \theta
\cos \left( {\psi -\Psi }\right) +g_{S}\cos \theta }\right\} -Tf_{S}\sin
\theta \sin \left( {\psi -\Psi }\right) }\right] ^{-1} \\
\mathrm{\ }\times \left\{ {\left[ {\ -g_{S}\sin \theta \sin \psi +\bar{f}%
_{S}\cos \theta \sin \Psi -f_{S}\cos \theta \cos \Psi }\right] \hat{\tau}_{1}%
}\right. \\
+\left[ {\ -\bar{f}_{S}\cos \theta \cos \Psi +g_{S}\sin \theta \cos \psi
-f_{S}\cos \theta \sin \Psi }\right] \hat{\tau}_{2} \\
\left. {\ +\left[ {\bar{f}_{S}\sin \theta \sin \left( {\psi -\Psi }\right)
+f_{S}\sin \theta \cos \left( {\psi -\Psi }\right) }\right] \hat{\tau}_{3}}%
\right\} ,%
\end{array}%
\end{equation}%
and hence we find the following form of the matrix current:
\begin{equation*}
\left\langle {\hat{I}_{R}}\right\rangle =i\left( {\begin{array}{*{20}c} { -
I_1 \sin \theta \sin \psi + I_2 \cos \theta \sin \Psi - I_3 \cos \theta \cos
\Psi } \\ { - I_2 \cos \theta \cos \Psi + I_1 \sin \theta \cos \psi - I_3
\cos \theta \sin \Psi } \\ {I_2 \sin \theta \sin \left( {\psi - \Psi }
\right) + I_3 \sin \theta \cos \left( {\psi - \Psi } \right)} \\ \end{array}}%
\right) \cdot \hat{\bm{\tau}},
\end{equation*}

\begin{equation*}
I_{1}=\left\langle {\frac{{2T_{m}g_{S}}}{{A}}}\right\rangle,
I_{2}=\left\langle {\frac{{2T_{m}\overline{f}_{S}}}{{A}}}\right\rangle,
I_{3}=\left\langle {\frac{{2T_{m}f_{S}}}{{A}}}\right\rangle,
\end{equation*}%
\begin{equation*}
A={{\left( {2 - T_{m}} \right) + T_{m}\left\{ {\bar f_S
\sin \theta \cos \left( {\psi - \Psi } \right) + g_S \cos \theta } \right\}
- T_{m} f_S \sin \theta \sin \left( {\psi - \Psi } \right)}},
\end{equation*}

\begin{equation}
g_S = \frac{{g_ + + g_ - }}{{1 + f_ + f_ - + g_ + g_ - }}, \bar f_S = \frac{{%
f_ + + f_ - }}{{1 + f_ + f_ - + g_ + g_ - }}, f_S = \frac{{i\left( {f_ + g_
- - f_ - g_ + } \right)}}{{1 + f_ + f_ - + g_ + g_ - }}.
\end{equation}

Finally the boundary conditions are given by
\begin{eqnarray}
\frac{{LR_b }}{{R_d }}\frac{\partial }{{\partial x}}\theta = - I_1 \sin
\theta + I_2 \cos \theta \cos \left( {\psi - \Psi } \right) - I_3 \cos
\theta \sin \left( {\psi - \Psi } \right), \\
\frac{{LR_b }}{{R_d }}\sin \theta \frac{\partial }{{\partial x}}\psi = -I_2
\sin\left( {\psi - \Psi } \right) - I_3 \cos \left( {\psi - \Psi } \right) .
\end{eqnarray}

For the calculation of the thermodynamical quantities, we use Matsubara
representation: $\varepsilon \rightarrow i\omega $. We parametrize the
quasiclassical Green's functions $G$ and $F$ using function $\Phi $:
\begin{eqnarray}
G_{\omega } &=&\frac{\omega }{\sqrt{\omega ^{2}+\Phi _{\omega }\Phi
_{-\omega }^{\ast }}}=\cos \theta , \\
F_{\omega } &=&\frac{{\Phi _{\omega }}}{\sqrt{\omega ^{2}+\Phi _{\omega
}\Phi _{-\omega }^{\ast }}}=\frac{{\Phi _{\omega }}}{\omega }G_{\omega
}=\sin \theta e^{-i\psi }, \\
F_{-\omega }^{\ast } &=&\frac{{\Phi _{-\omega }^{\ast }}}{\sqrt{\omega
^{2}+\Phi _{\omega }\Phi _{-\omega }^{\ast }}}=\frac{{\Phi _{-\omega }^{\ast
}}}{\omega }G_{\omega }=\sin \theta e^{i\psi }
\end{eqnarray}%
with Matsubara frequency $\omega $. Then Usadel equation reads\cite{Usadel}
\begin{equation}
\xi ^{2}\frac{{\pi T_{C}}}{{G_{\omega }}}\frac{\partial }{{\partial x}}%
\left( {G_{\omega }^{2}\frac{\partial }{{\partial x}}\Phi _{\omega }}\right)
-\left( {\omega -(+)ih}\right) \Phi _{\omega }=0
\end{equation}%
for majority (minority) spin with $\xi =\sqrt{D/2\pi T_{C}}$ and critical
temperature $T_{C}$. The following relations are satisfied:
\begin{eqnarray}
\sin \theta \cos \psi &=&\frac{{G_{\omega }}}{{2\omega }}\left( {\Phi
_{\omega }+\Phi _{-\omega }^{\ast }}\right) , \\
\sin \theta \sin \psi &=&\frac{{iG_{\omega }}}{{2\omega }}\left( {\Phi
_{\omega }-\Phi _{-\omega }^{\ast }}\right) .
\end{eqnarray}

Then the boundary condition is expressed as
\begin{equation*}
\frac{{G_{\omega }}}{\omega }\frac{\partial }{{\partial x}}\Phi _{\omega }=
\frac{{R_{d}}}{{R_{b}L}}\left( {-\frac{{\Phi _{\omega }}}{\omega }%
I_{1}+e^{-i\Psi }\left( {I_{2}+iI_{3}}\right) }\right)
\end{equation*}%
\begin{equation*}
I_{1}=\left\langle {\frac{{2T_{m}g_{S}}}{{A}}}\right\rangle,
I_{2}=\left\langle {\frac{{2T_{m}\overline{f}_{S}}}{{A}}}\right\rangle,
I_{3}=\left\langle {\frac{{2T_{m}f_{S}}}{{A}}}\right\rangle,
\end{equation*}%
\begin{equation*}
A=2-T_{m}+T_{m}(g_{S}G_{\omega } +\overline{f}_{S}(B\cos \Psi +C\sin \Psi
)-f_{S}(C\cos \Psi -B\sin \Psi )),
\end{equation*}%
\begin{equation}
B=\frac{{G_{\omega }}}{{2\omega }}\left( {\Phi _{\omega }+\Phi _{-\omega
}^{\ast }}\right), \quad C=\frac{{iG_{\omega }}}{{2\omega }}\left( {\Phi
_{\omega }-\Phi _{-\omega }^{\ast }}\right)
\end{equation}
at $x=L$.

This boundary condition is quite general since with a proper choice of
$\Delta _{\pm }$, it is applicable to any unconventional superconductor with $%
S_{z}=0$ in a time reversal symmetry conserving state. Here, $S_{z}$ denotes
the $z$-component of the total spin of a Cooper pair. For $s$-, $d$- and $p$%
-wave superconductors we choose $\Delta _{\pm }=\Delta (T),\Delta (T)\cos
(2\phi \mp 2\alpha )$ and $\Delta (T)\cos (\phi \mp \alpha )$ respectively.

In the following we will calculate Josephson current using this boundary
condition at $x=0$ and $x=L$, where $\varphi $ is the external phase
difference across the junctions, and $\alpha $ and $\beta $ denote the angles
between the normal to the interface and the crystal axes of USCs for $x\leq
0 $ and $x\geq L$ respectively. It is important to note that the solution of
the Usadel equation is invariant under the transformation $\alpha
\rightarrow -\alpha $ or $\beta \rightarrow -\beta $. This is clear by
replacing $\phi $ with $-\phi $ in the angular averaging.

%%%%%%%%%%%%%%%%%%%%%%%%%%%%%%%%%%%%%%%%%%%%%%%%%%%%%%%%%%%%%%%%%%%%%%%%%%
%  Actual calculation to obtain electric current
%%%%%%%%%%%%%%%%%%%%%%%%%%%%%%%%%%%%%%%%%%%%%%%%%%%%%%%%%%%%%%%%%%%%%%%%%%
Josephson current is given by the expression
\begin{equation}
\frac{{eIR}}{{\pi T_C }} = i\frac{{RTL}}{{4R_d T_C }}\sum\limits_{ \uparrow
, \downarrow ,\omega } {\frac{{G_\omega ^2 }}{{\omega ^2 }}} \left( {\Phi
_\omega \frac{\partial }{{\partial x}}\Phi _{ - \omega }^ * - \Phi _{ -
\omega }^ * \frac{\partial }{{\partial x}}\Phi _\omega } \right)
\end{equation}
with temperature $T$ and $R \equiv R_{d}+R_{b}+R_{b}^{\prime}$. Below $I_C$
denotes the critical current and we consider symmetric barriers with $%
R_{b}=R^{\prime }_{b}$ and $Z=Z^{\prime }$ for simplicity.

%It is important to note that in the present approach, according to
%the circuit theory, $R_{d}/R_{b}^{(\prime )}$ can be varied
%independently of $ T_{m}^{(\prime )}$, $i.e.$, independently of
%$Z^{(\prime )}$ (see Ref. \cite {TGK}), thus $R_{d}/R_{b}^{(\prime
%)}$ and $Z^{(\prime )}$ can be chosen as independent parameters.

%%%%%%%%%%%%%%%%%%%%%%%%%%%%%%%
% CHANGE
%%%%%%%%%%%%%%%%%%%%%%%%%%%%%%%%

\section{Results}

\subsection{Junctions with DN}

\begin{figure}[htb]
\begin{center}
\scalebox{0.4}{
\includegraphics[width=16.0cm,clip]{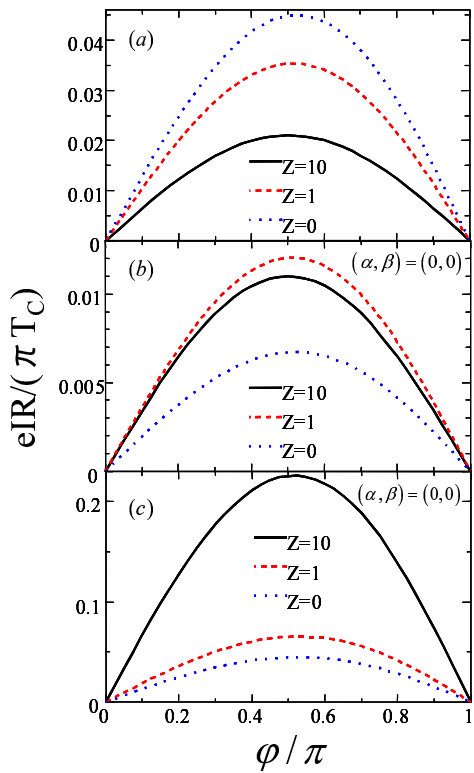}}
\end{center}
\caption{ (color online) Current-phase relation for $T/T_{C}=0.1$, $%
R_d/R_b=0.1$ and $E_{Th}/\Delta(0)=0.2$. (a)$s$-wave junctions. (b)$d$-wave
junctions. (c)$p$-wave junctions. }
\label{f1}
\end{figure}

\begin{figure}[htb]
\begin{center}
\scalebox{0.4}{
\includegraphics[width=16.0cm,clip]{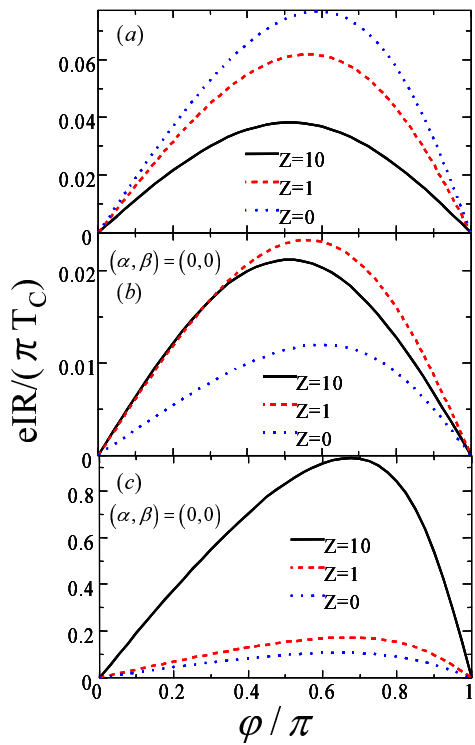}}
\end{center}
\caption{(color online) Current-phase relation for $T/T_{C}=0.02$, $%
R_d/R_b=0.1$ and $E_{Th}/\Delta(0)=0.2$. (a)$s$-wave junctions. (b)$d$-wave
junctions. (c)$p$-wave junctions. }
\label{f2}
\end{figure}

Let us first focus on the junctions with DN. Figure \ref{f1} shows the
current-phase relation for $T/T_{C}=0.1$, $R_{d}/R_{b}=0.1$ and $E_{Th}/\Delta
(0)=0.2$ in (a) $s$-wave, (b)$d$-wave and (c)$p$-wave superconducting
junctions with $\left( {\alpha ,\beta }\right) =\left( {0,0}\right) $. In $s$%
-wave junctions, the $IR$ product is suppressed with the increase of $Z$
because proximity effect is suppressed. In $d$-wave junctions, proximity
effect and hence $IR$ are enhanced with the increase of $Z$ because of the
cancellation of the positive and negative parts of pair potential in the
angular averaging. As $Z$ increases, the contribution from the positive part
exceeds that from negative part and hence the cancellation becomes weak\cite%
{Yoko}. In $p$-wave junctions, the $IR$ product is strongly enhanced with
the increase of $Z$ because of the formation of the resonant states. It is
known that the proximity effect and MARS can coexist \cite{p-wave}. At $%
\left( {\alpha ,\beta }\right) =\left( {0,0}\right) $, proximity effect is
mostly enhanced. As $Z$ increases, the contribution of the MARS becomes
remarkable and hence the proximity effect gets strongly enhanced\cite{p-wave}%
. Consequently its magnitude is an order of magnitude larger than that in $s$%
-wave junctions. The results at lower temperatures are shown in Fig. \ref{f2}%
. The $IR$ product is enhanced with decreasing temperature because the
proximity effect is enhanced. In this case the current-phase relation in $p$%
-wave junctions has the form close to $\sin \varphi /2$, in contrast to the
standard sinusoidal relation. This is a peculiar property of the formation
of the resonant states in $p$-wave junctions where constructive interference
occurs near $\varphi =\pi $. \cite{AsanoL}

\begin{figure}[htb]
\begin{center}
\scalebox{0.4}{
\includegraphics[width=25.0cm,clip]{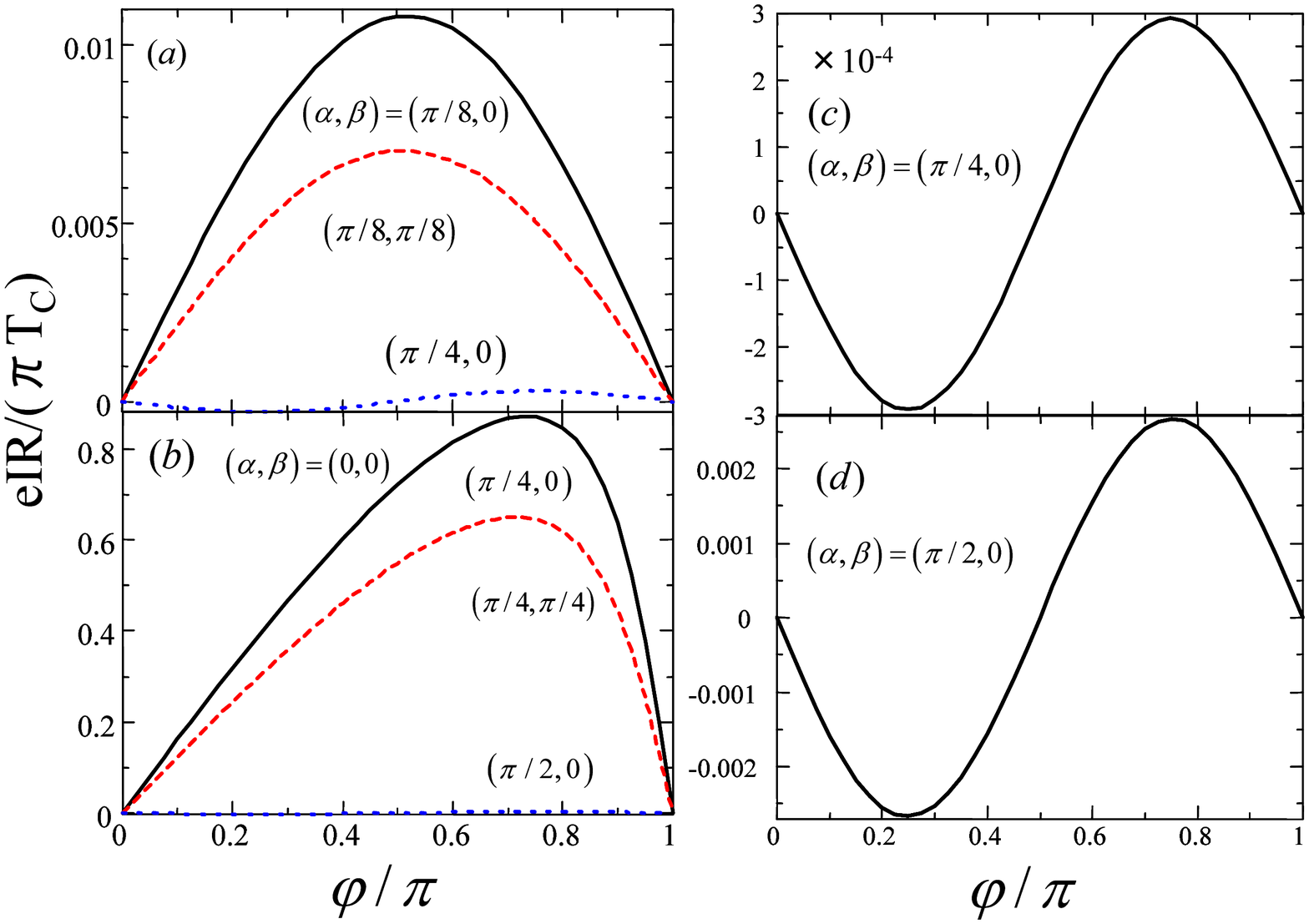}}
\end{center}
\caption{(color online) Current-phase relation for $T/T_{C}=0.01$, $Z=10$, $%
R_d/R_b=0.1$ and $E_{Th}/\Delta(0)=0.2$.(a) and (c)$d$-wave junctions. (b)
and (d)$p$-wave junctions. }
\label{f3}
\end{figure}

Next we study the Josephson effect for other misorientational angles. As $%
\alpha $ or $\beta $ increase, the $IR$ product is monotonically suppressed
due to the suppression of the proximity effect as shown in Fig. \ref{f3}(a)
and Fig. \ref{f3}(b). In $d(p)$-wave junctions, the first harmonics
disappear and hence the $IR$ product is proportional to $-\sin 2\varphi $ at
$\left( {\alpha ,\beta }\right) =\left( {\pi /4,0}\right) (\left( {\pi /2,0}%
\right) )$ as shown in Figs. \ref{f3} (c) and (d). This can be explained in
the limiting case as follows. Near $T_{C}$, the $IR$ product, which stems
from the first harmonics, is proportional to $\cos 2\alpha \cos 2\beta $ in $%
d$-wave junctions because angular averaging gives $<\cos (2\phi -2\alpha
)>\propto \cos 2\alpha $.\cite{Yoko2} Thus the first harmonics disappear at $%
\alpha =\pi /4$. Similar argument is also applicable to $p$-wave junctions.

\begin{figure}[tbh]
\begin{center}
\scalebox{0.4}{
\includegraphics[width=16.0cm,clip]{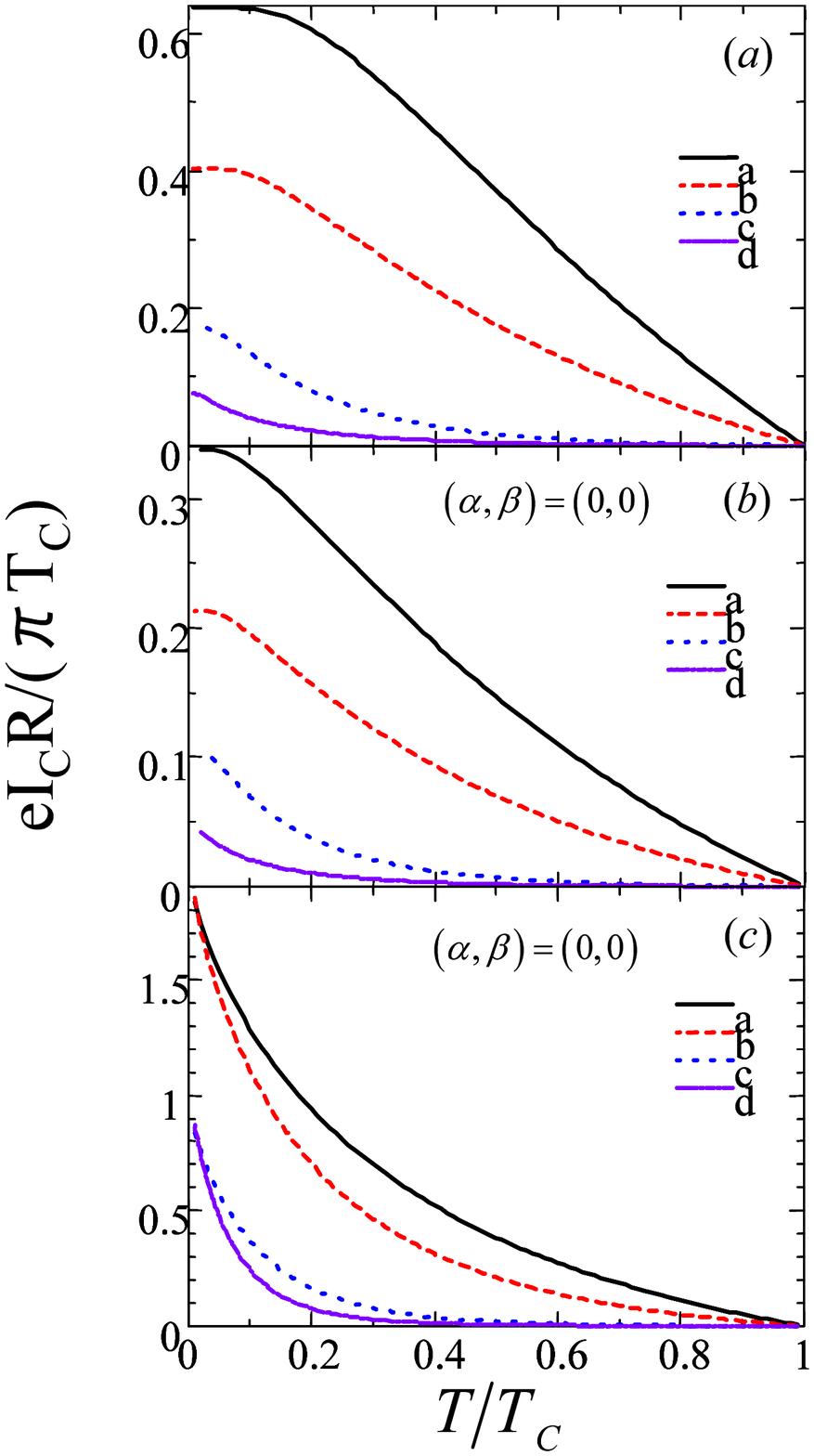}}
\end{center}
\caption{(color online) Temperature dependence of the critical current. (a)$%
s $-wave junctions. (b)$d$-wave junctions. (c)$p$-wave junctions. a.$%
R_{d}/R_{b}=2$ and $E_{Th}/\Delta (0)=1$. b.$R_{d}/R_{b}=0.5$ and $%
E_{Th}/\Delta (0)=1$. c.$R_{d}/R_{b}=2$ and $E_{Th}/\Delta (0)=0.1$. d.$%
R_{d}/R_{b}=0.5$ and $E_{Th}/\Delta (0)=0.1$.}
\label{f4}
\end{figure}

\begin{figure}[htb]
\begin{center}
\scalebox{0.4}{
\includegraphics[width=16.0cm,clip]{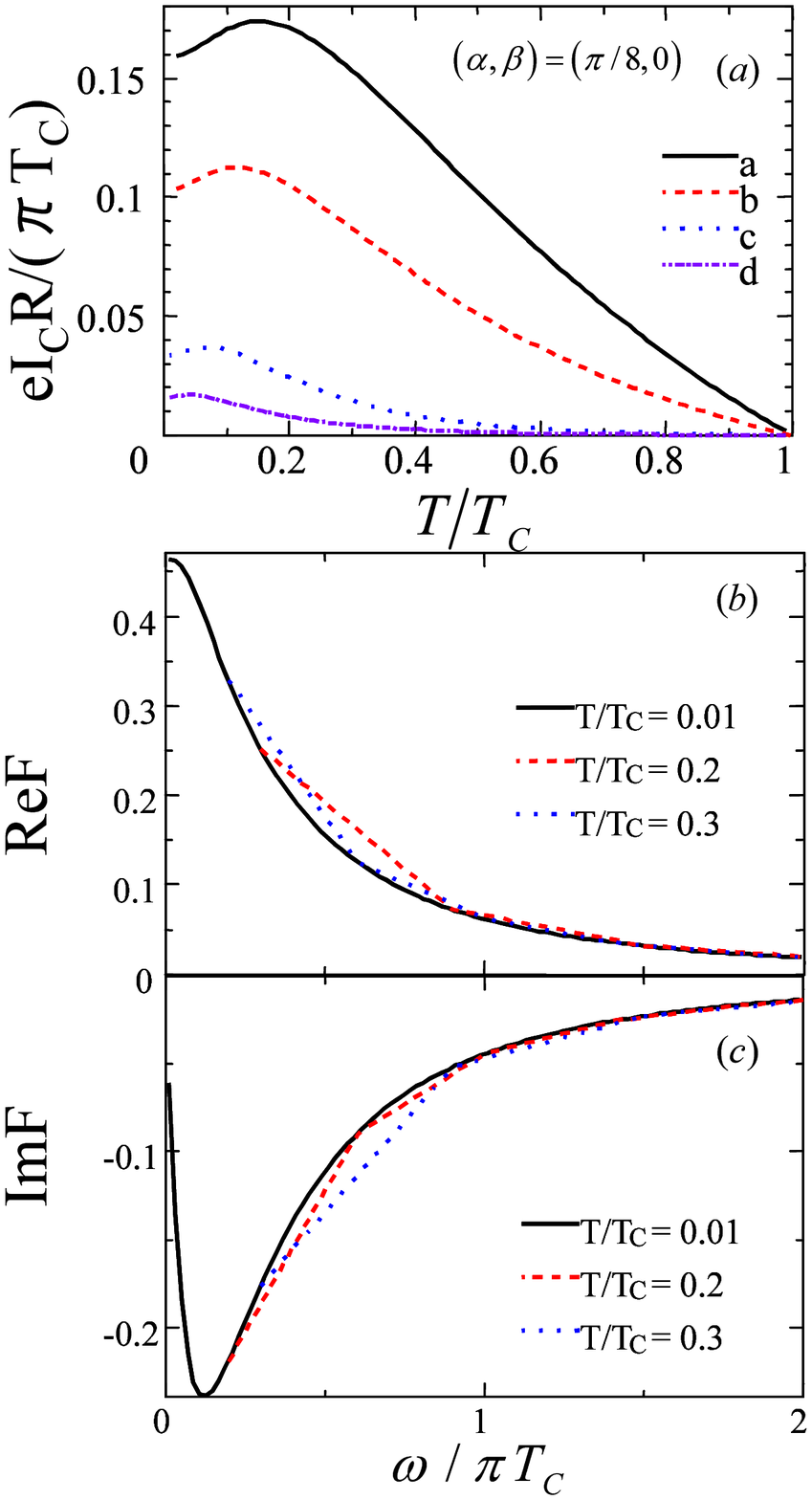}}
\end{center}
\caption{(color online) (a) Temperature dependence of the critical current
for $\left( {\protect\alpha ,\protect\beta } \right) = \left( {\protect\pi%
/8,0} \right)$. a.$R_d/R_b=2$ and $E_{Th}/\Delta(0)=1$. b.$R_d/R_b=0.5$ and $%
E_{Th}/\Delta(0)=1$. c.$R_d/R_b=2$ and $E_{Th}/\Delta(0)=0.1$. d.$%
R_d/R_b=0.5 $ and $E_{Th}/\Delta(0)=0.1$. (b) real and (c) imaginary parts
of anomalous Green's functions $F$ with $R_d/R_b=2$ and $E_{Th}/\Delta(0)=1$%
. }
\label{f5}
\end{figure}

Let us discuss the results for the critical current. In Fig. \ref{f4},
temperature dependence of the critical current is plotted for various $%
R_{d}/R_{b}$ and $E_{Th}/\Delta (0)$ in (a) $s$-wave, (b) $d$-wave and (c) $%
p $-wave superconducting junctions with $Z=10$ and $\left( {\alpha ,\beta }%
\right) =\left( {0,0}\right) $. As $R_{d}/R_{b}$ and $E_{Th}/\Delta (0)$
increase, $I_{C}R$ increases for all the junctions because proximity effect
is enhanced. In $p$-wave junctions, the critical current is strongly
enhanced at low temperatures compared to the $s$-wave and $d$-wave
junctions. When the misorientational angles in $d$-wave junctions are changed,
nonmonotonic temperature dependence appears as shown in Fig. \ref{f5} (a).
This nonmonotonic behavior can be explained in terms of the competition
between the proximity effect and the formation of MARS. It is known from the
previous studies that for $\alpha =\beta =0$ the proximity effect exists but
MARS is absent at the interfaces. On the other hand, for $\alpha =\beta =\pi
/4$, only MARS exists and the proximity effect is absent \cite%
{Nazarov2003,TNGK}. In other cases, both the proximity effect and MARS are
present. With the decrease of temperature, the formation of MARS strongly
suppresses the proximity effect. This results in the suppression of the
Josephson current at low temperatures. Therefore, a nonmonotonic temperature
dependence appears when both the proximity effect and MARS coexist.

The above statement can be confirmed by calculation of the dependence of
anomalous Green's function $F$ on Matsubara frequency $\omega $ as shown in
Figs. \ref{f5} (b) and \ref{f5} (c) at $x=L/2$ and $\varphi =\pi /2$ for $%
\left( {\alpha ,\beta }\right) =\left( {\pi /8,0}\right) $. At low
temperature ($T/T_{C}=0.01$) the magnitude of Im$F$ is suppressed at low
energy in contrast to the case of high temperature ($T/T_{C}=0.2$ and $0.3$%
). This result illustrates strong suppression of the proximity effect by the
formation of MARS at low $T$, which leads to the nonmonotonic temperature
dependence. Note that this nonmonotonic dependence can appear only for large
$Z$ when the role of MARS is essential.\cite{Yoko}

\begin{figure}[tbh]
\begin{center}
\scalebox{0.4}{
\includegraphics[width=16.0cm,clip]{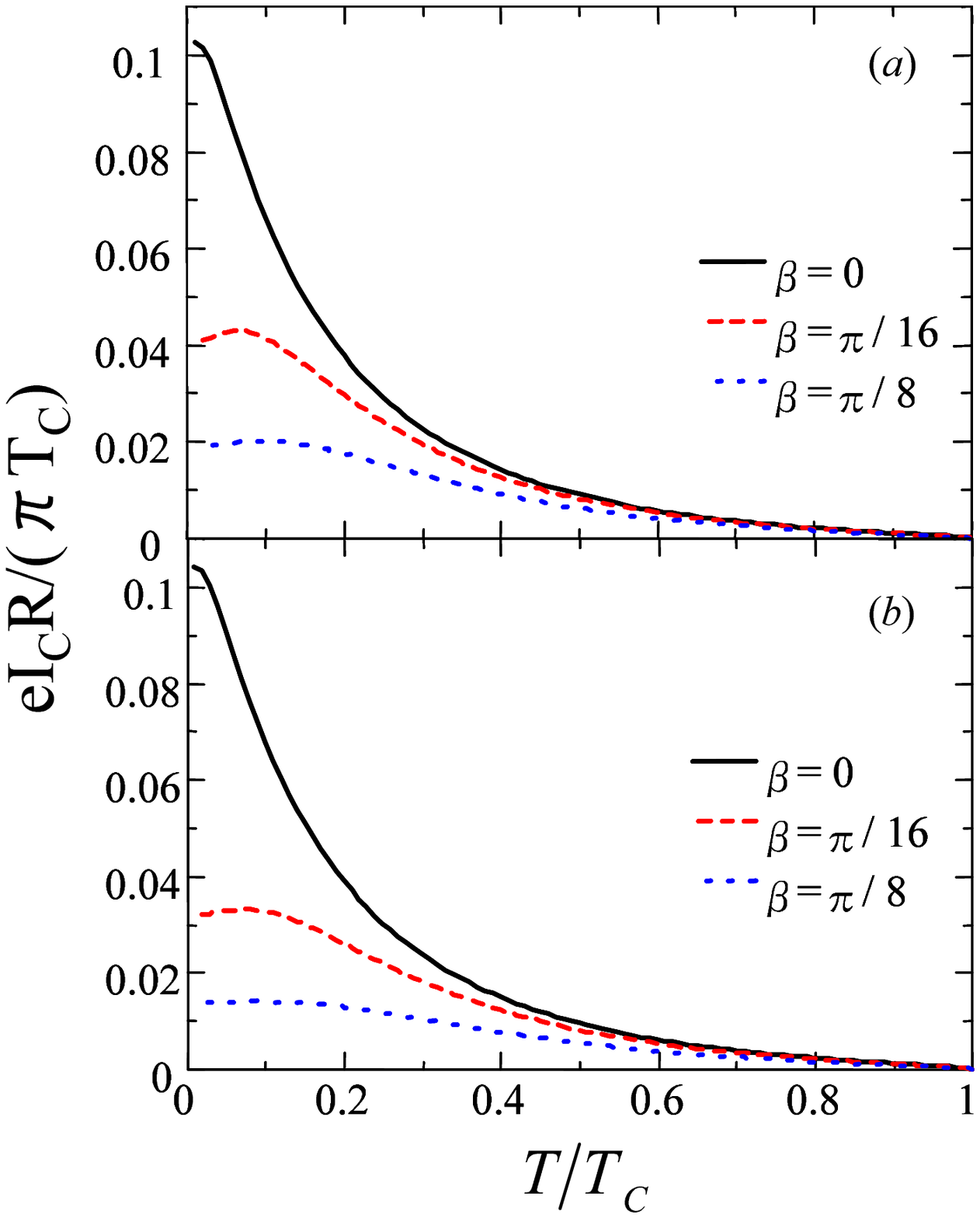}}
\end{center}
\caption{Temperature dependence of the critical current in S/DN/D junctions
with $Z=10$, $R_{d}/R_{b}=1$ and $E_{Th}/\Delta (0)=0.1$. (a) $%
T_{C}/T_{CS}=5 $ and (b) $T_{C}/T_{CS}=10$. }
\label{f6}
\end{figure}

It is interesting to study the junctions composed of superconductors with
different symmetries. Here we study S/DN/D junctions with $Z=10$, $%
R_{d}/R_{b}=1$ and $E_{Th}/\Delta (0)=0.1$. We choose $T_{CD}/T_{CS}=5$ in
Fig. \ref{f6}(a) and $T_{CD}/T_{CS}=10$ in Fig. \ref{f6}(b) where $%
T_{CS}(T_{CD}$) denotes the critical temperature of the $s$-wave ($d$-wave)
superconductors. In this case the nonmonotonic temperature dependence also
occurs due to the competition as shown in Fig. \ref{f6}. In S/I/D junctions,
the nonmonotonic temperature dependence was observed experimentally in Ref.
\cite{Iguchi}. We can qualitatively explain these data by regarding the
barrier as a diffusive normally conducting material.

\begin{figure}[htb]
\begin{center}
\scalebox{0.4}{
\includegraphics[width=16.0cm,clip]{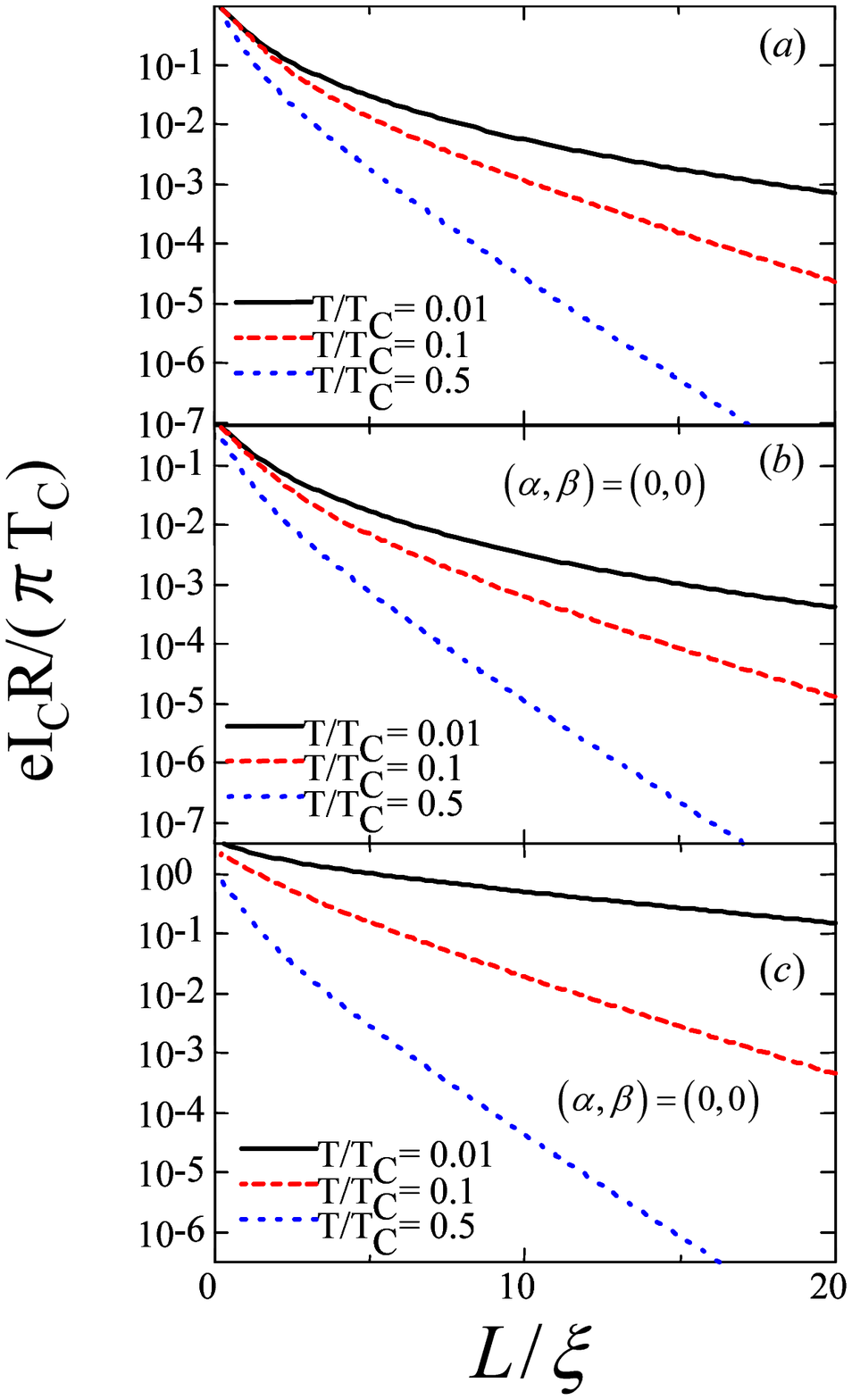}}
\end{center}
\caption{(color online) Length dependence of the critical current with $Z=10$ 
 and $R_d/R_b=0.1$, (a)$s$-wave junctions. (b)$d$-wave junctions. (c)$p$-wave
junctions.}
\label{f7}
\end{figure}

\begin{figure}[htb]
\begin{center}
\scalebox{0.4}{
\includegraphics[width=16.0cm,clip]{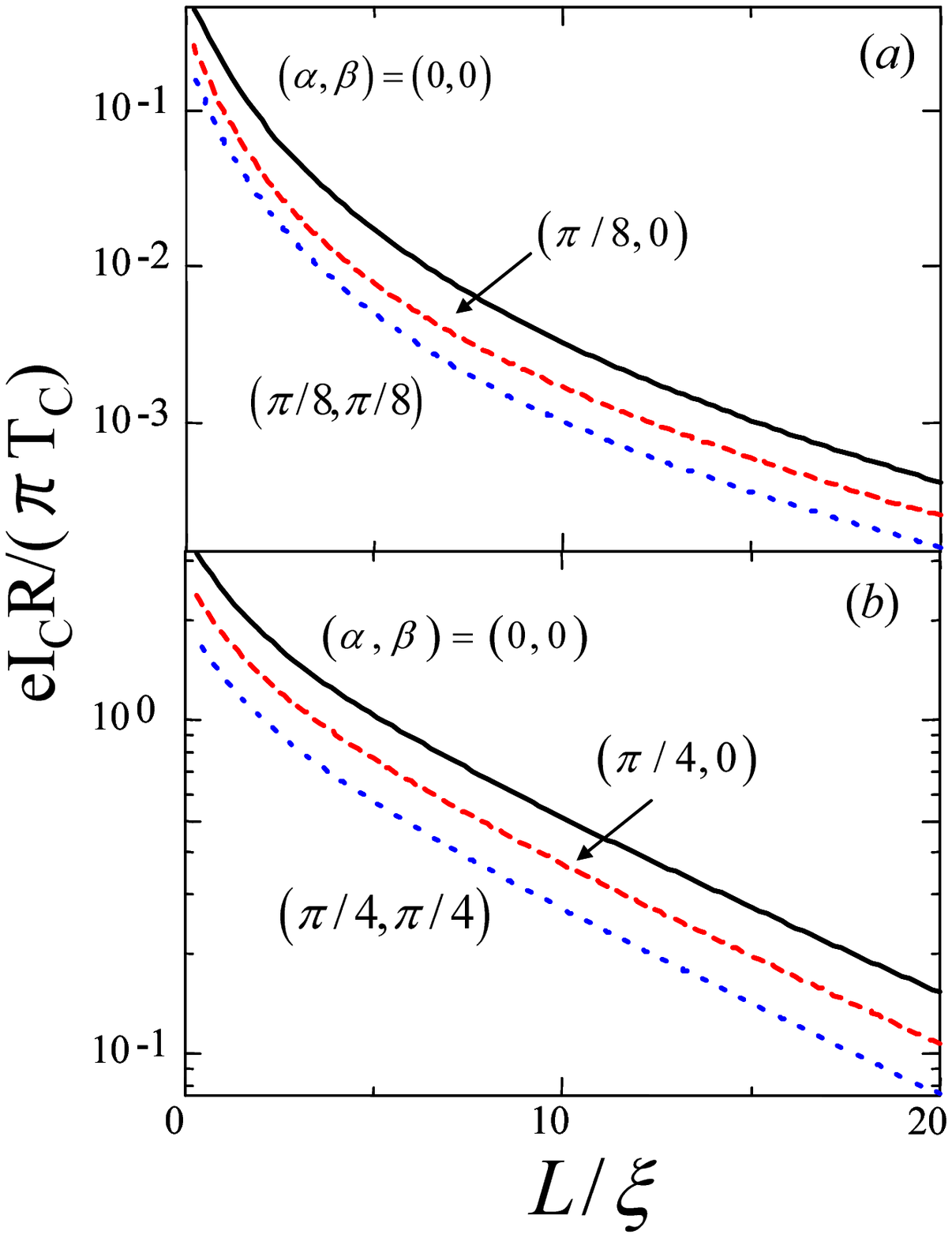}}
\end{center}
\caption{(color online) Length dependence of the critical current with $%
T/T_C=0.01$, $Z=10$ and $R_d/R_b=0.1$. (a)$d$-wave junctions. (b)$p$-wave
junctions.}
\label{f8}
\end{figure}

We study dependence of the critical current on barrier thickness $L$ at
various temperatures in (a) $s$-wave, (b)$d$-wave and (c)$p$-wave
superconducting junctions with $Z=10$, $R_{d}/R_{b}=0.1$ and $\left( {\alpha
,\beta }\right) =\left( {0,0}\right) $ in Fig. \ref{f7}. The $I_{C}R$
product is proportional to $exp(-\bar{C}L/\xi )$ for large $L/\xi $ for all
the junctions as shown in Fig. \ref{f7}. Here $\bar{C}$ is a constant
independent of $L$. As temperature is lowered, the magnitude of $\bar{C}$ is
reduced. From our results, we also find the relation $\bar{C}\propto
T^{-1/2} $. The results for the junctions with other misorientational angles
for $T/T_{C}=0.01$, $Z=10$ and $R_{d}/R_{b}=0.1$ are shown in Fig. \ref{f8}.
As $\alpha $ and $\beta $ increase, $I_{C}R$ is suppressed. However, $\bar{C}
$ is independent of these values, which indicates that MARS don't influence
the effective coherence length $\xi /\bar{C}$. This is because the effective
coherence length reflecting the penetration of Cooper pairs is determined by
the Usadel equation and therefore is independent of MARS.

\subsection{Junctions with DF}

\begin{figure}[htb]
\begin{center}
\scalebox{0.4}{
\includegraphics[width=16.0cm,clip]{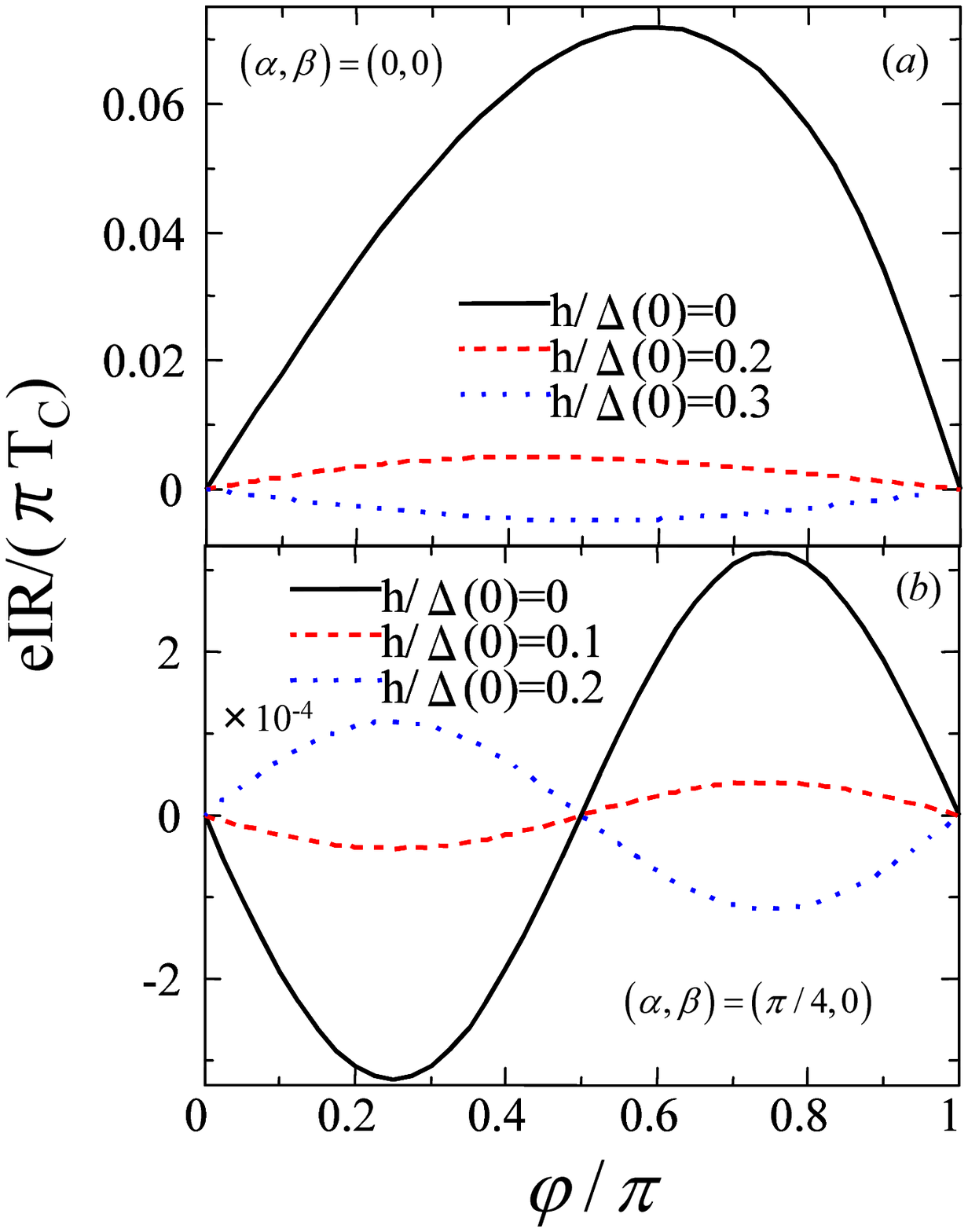}}
\end{center}
\caption{ (color online) Current-phase relation in D/DF/D junctions for $%
T/T_{C}=0.01$, $Z=10$, $R_d/R_b=1$ and $E_{Th}/\Delta(0)=0.1$. }
\label{f9}
\end{figure}

\begin{figure}[htb]
\begin{center}
\scalebox{0.4}{
\includegraphics[width=26.0cm,clip]{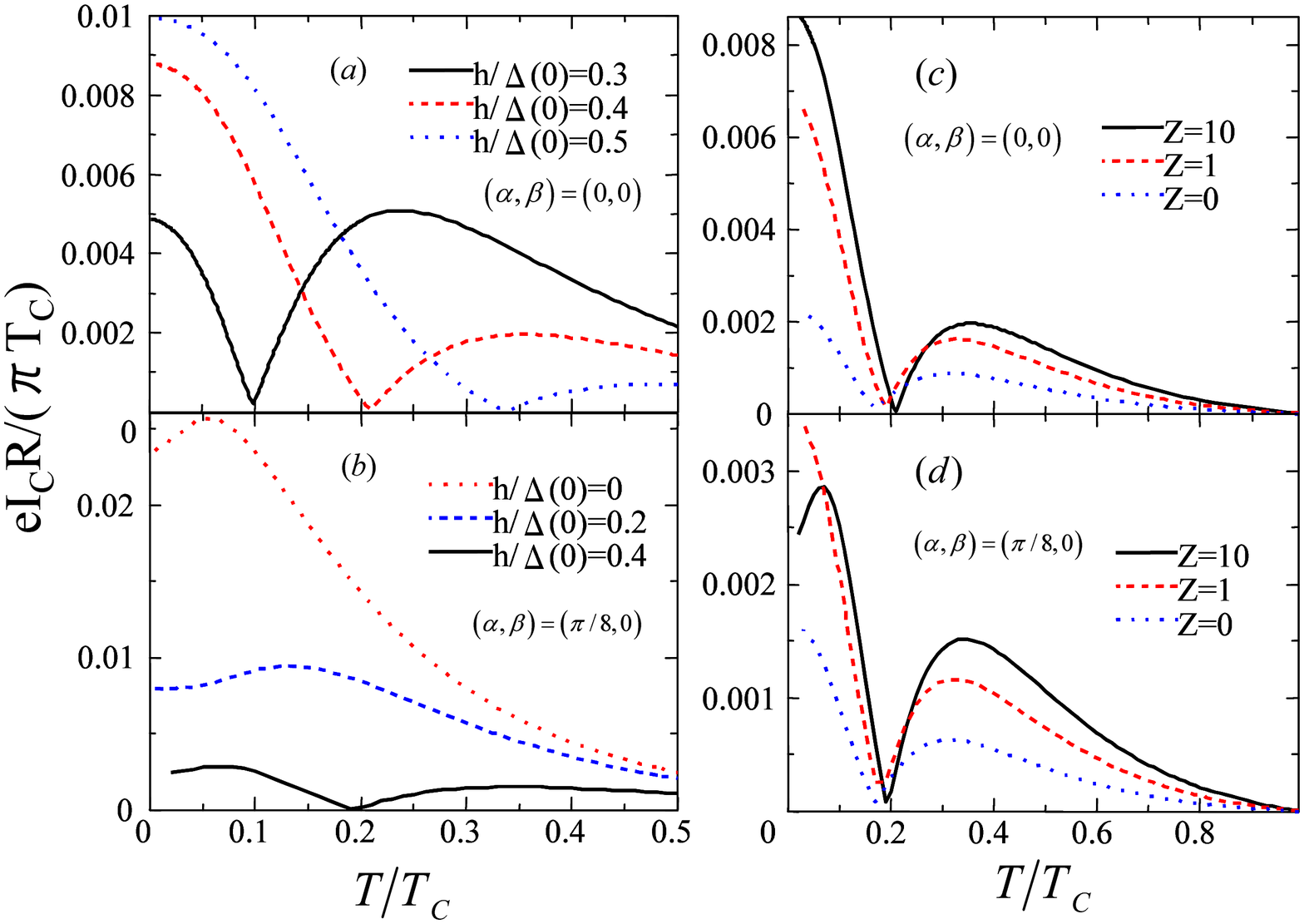}}
\end{center}
\caption{(color online) Temperature dependence of the critical current in
D/DF/D junctions. $Z=10$ in (a) and (b). $h/\Delta(0)=0.4$ in (c) and (d). }
\label{f10}
\end{figure}

Here we consider junctions with DF. We will study three types of junctions:
D/DF/D, P/DF/P and S/DF/P junctions. Figure \ref{f9} shows current-phase
relation in D/DF/D junctions for $T/T_{C}=0.01$, $Z=10$, $%
R_{d}/R_{b}=1$ and $E_{Th}/\Delta (0)=0.1$. At $\left( {\alpha ,\beta }\right)
=\left( {0,0}\right) $ where the MARS are absent, the exchange field causes
a 0-$\pi $ transition as predicted for $s$-wave junctions (see Fig. \ref{f9}
(a)). Similarly, second harmonic changes its sign at $\left( {\alpha ,\beta }%
\right) =\left( {\pi /4,0}\right) $, where the proximity effect is absent at
$x=0$, as shown in Fig. \ref{f9} (b). Figure \ref{f10} displays temperature
dependence of the critical current in D/DF/D junctions with $%
R_{d}/R_{b}=1$ and $E_{Th}/\Delta (0)=0.1$. At $\left( {\alpha ,\beta }\right)
=\left( {0,0}\right) $, the exchange field causes a 0-$\pi $ transition as
shown in Fig. \ref{f10} (a). At $\left( {\alpha ,\beta }\right) =\left( {\pi
/8,0}\right) $, the exchange field also causes a 0-$\pi $ transition, and as
a result, double peak structure appears for $h/\Delta (0)=0.4$ as shown in
Fig. \ref{f10} (b). The peak at lower temperature stems from the competition
between proximity effect and MARS. The peak at higher temperature stems from
the 0-$\pi $ transition. With the decrease of $Z$, the magnitude of $I_{C}R$
is suppressed while the 0-$\pi $ transition temperature is almost
independent of $Z$ (see Fig. \ref{f10} (c)). At $\left( {\alpha ,\beta }%
\right) =\left( {\pi /8,0}\right) $, the peak at lower temperature
disappears for small $Z$ as shown in Fig. \ref{f10} (d) because the
existence of the insulating barrier is essential for the formation of MARS.

\begin{figure}[htb]
\begin{center}
\scalebox{0.4}{
\includegraphics[width=16.0cm,clip]{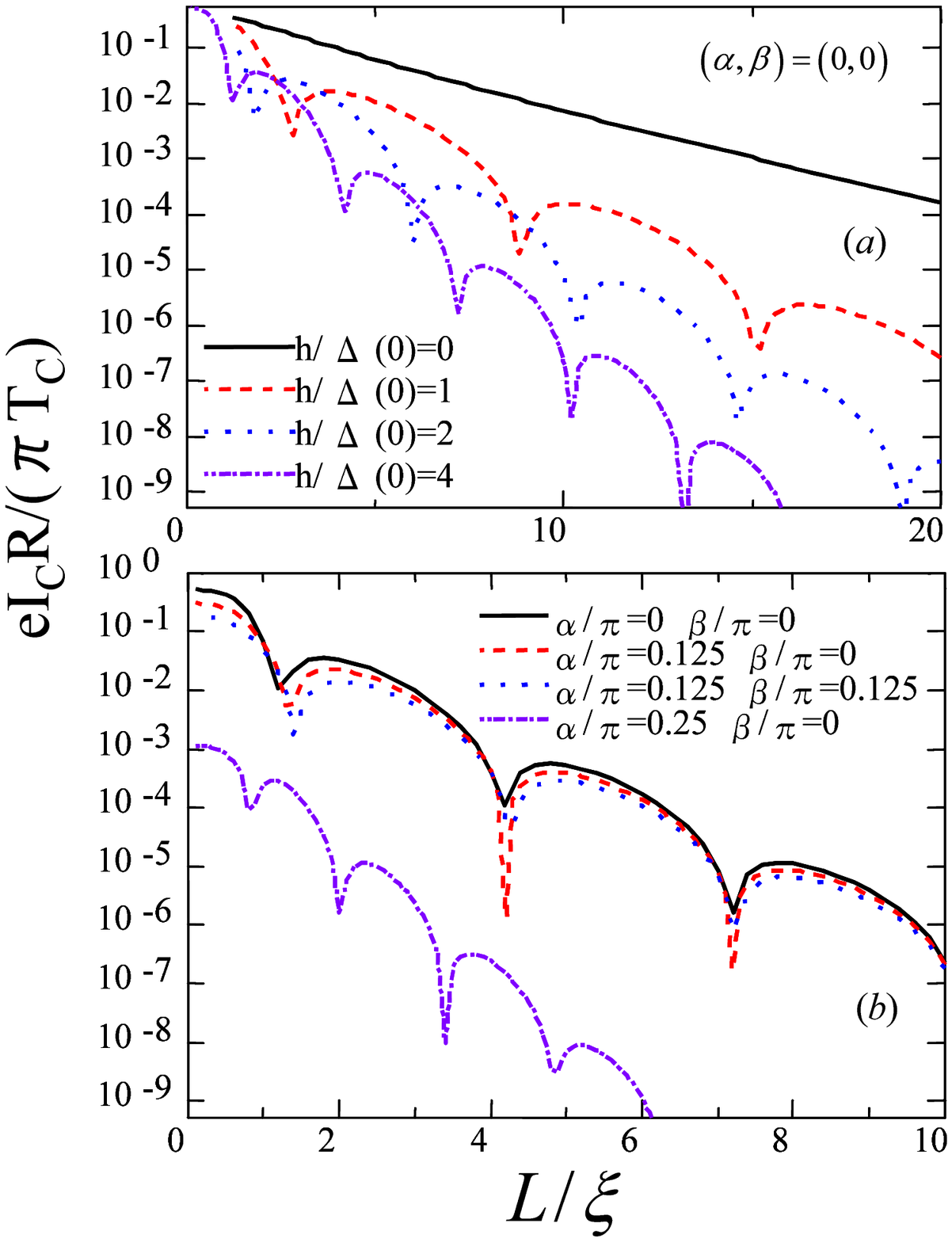}}
\end{center}
\caption{(color online) Length dependence of the critical current in D/DF/D
junctions with $T/T_{C}=0.1$, $Z=10$, $R_d/R_b=1$ and $E_{Th}/\Delta(0)=0.1$%
. We choose $h/\Delta(0)=4$ in (b).}
\label{f11}
\end{figure}

The barrier thickness dependence of the critical current in D/DF/D junctions
is plotted in Fig.\ref{f11} with $T/T_{C}=0.1$, $Z=10$, $%
R_{d}/R_{b}=1$ and $E_{Th}/\Delta (0)=0.1$. For $h=0$, the $I_{C}R$ product has an
exponential dependence on $L$. As $h/\Delta (0)$ increases, $I_{C}$
oscillates as a function of $L/\xi $. The period of the oscillation becomes
 shorter with increasing $h$ as shown in Fig.\ref{f11} (a). As $\alpha $ and $%
\beta $ increase, $I_{C}R$ is suppressed while the period of the
oscillations remains constant, that is, the period is independent of the
MARS. The second harmonics have a shorter (almost half) oscillation period
than that of the first harmonics, similar to the predictions for S/DF/S
junctions\cite{Buzdin3,Houzet} (see the result for $\left( {\alpha ,\beta }%
\right) =\left( {\pi /4,0}\right) $ in Fig.\ref{f11} (b)).

\begin{figure}[htb]
\begin{center}
\scalebox{0.4}{
\includegraphics[width=16.0cm,clip]{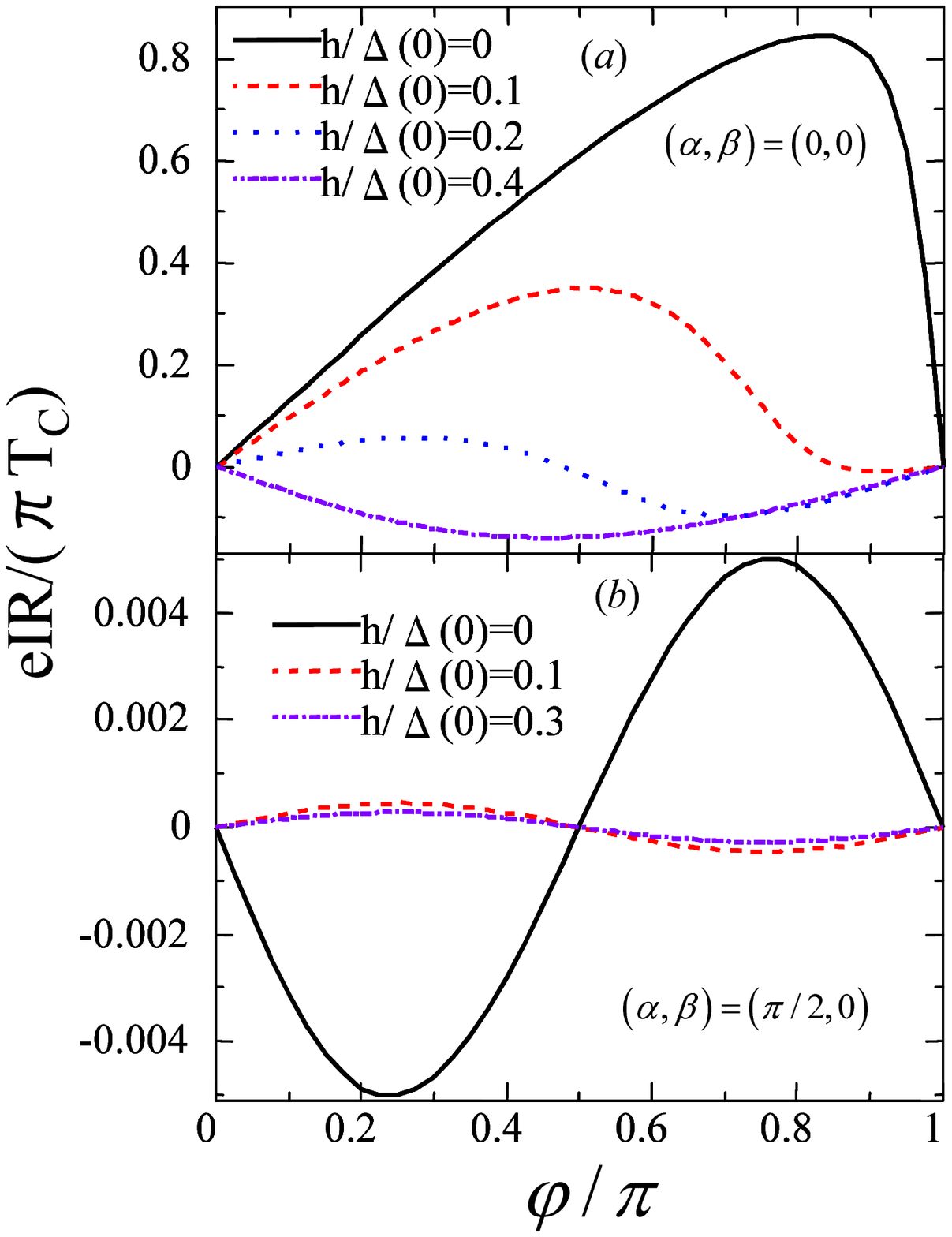}}
\end{center}
\caption{ (color online) Current-phase relation in P/DF/P junctions for $%
T/T_{C}=0.01$, $Z=10$, $R_d/R_b=1$ and $E_{Th}/\Delta(0)=0.1$. }
\label{f12}
\end{figure}

\begin{figure}[htb]
\begin{center}
\scalebox{0.4}{
\includegraphics[width=16.0cm,clip]{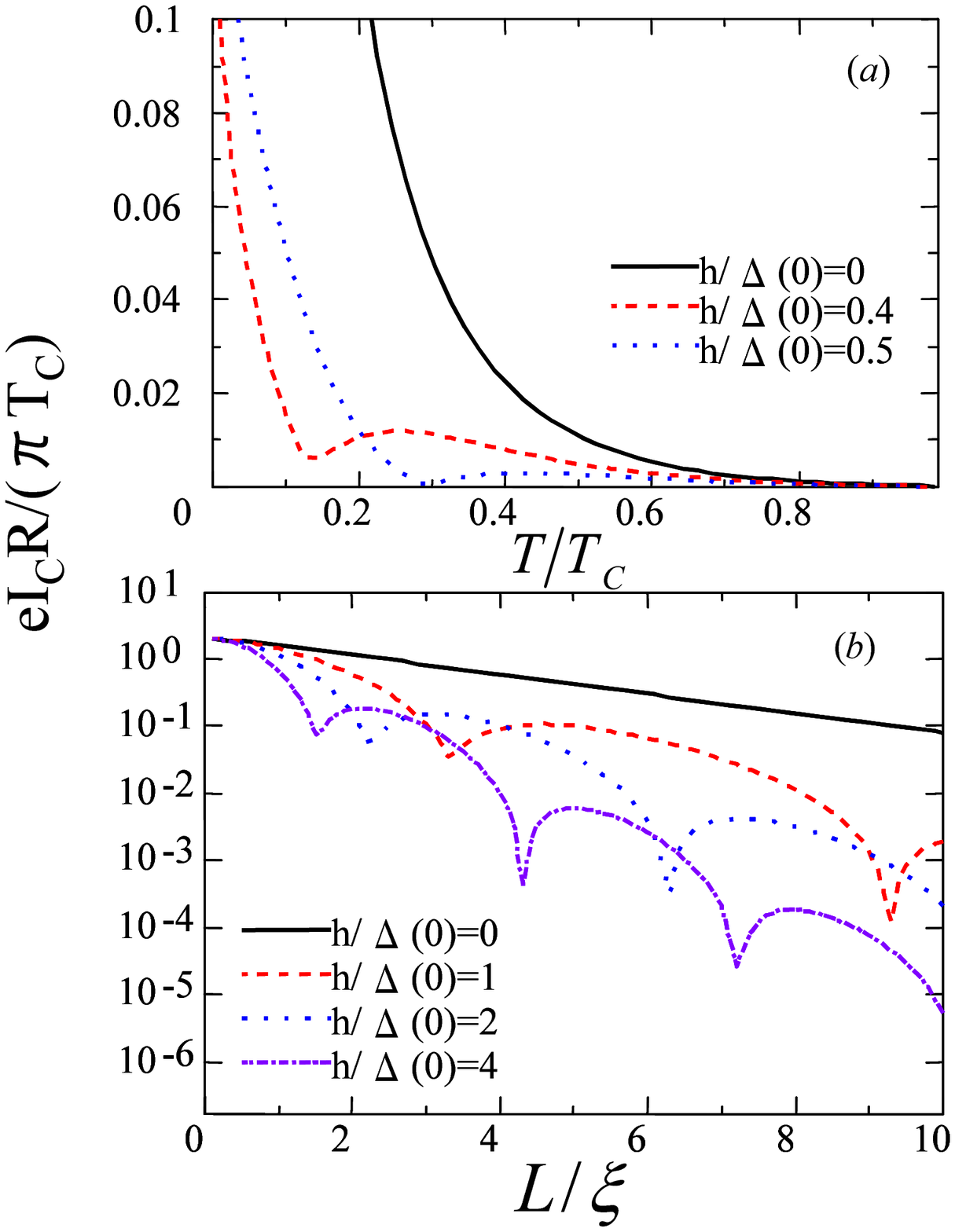}}
\end{center}
\caption{(color online) (a) temperature and (b) length dependence of the
critical current in P/DF/P junctions. $Z=10$, $R_d/R_b=1$, and $\left( {%
\protect\alpha ,\protect\beta } \right) = \left( {0,0} \right)$. We choose $%
E_{Th}/\Delta(0)=0.1$ in (a) and $T/T_{C}=0.1$ in (b). }
\label{f13}
\end{figure}

Next we consider the P/DF/P junctions. Current-phase relation in P/DF/P
junctions for $T/T_{C}=0.01$, $Z=10$, $R_{d}/R_{b}=1$ and $E_{Th}/\Delta
(0)=0.1$ is plotted in Fig.\ref{f12}. With increasing $h$, the dependence
of $IR$ changes from $\sin \varphi /2$ to $\sin 2\varphi $ and finally to $%
-\sin \varphi $ at $\left( {\alpha ,\beta }\right) =\left( {0,0}\right) $ as
shown in Fig.\ref{f12} (a). The phase dependences originate from the
formation of the resonant states, the disappearance of the first harmonics at the 0-$\pi$ transition 
and the emergence of the $\pi$-junctions, respectively. At $\left( {\alpha
,\beta }\right) =\left( {\pi /2,0}\right) $ where the MARS are absent at $x=0
$, the second harmonics change the sign with the increase of $h$ as shown in
Fig.\ref{f12} (b). The critical current as a function of $T$ is shown in Fig.%
\ref{f13} (a). The 0-$\pi$ transition occurs due to the exchange field.
Similarly, as $h/\Delta (0)$ increases, $I_{C}R$ oscillates as a function of
$L/\xi $. The period of the oscillation becomes short with increasing $h$ as
shown in Fig.\ref{f13} (b).

\begin{figure}[htb]
\begin{center}
\scalebox{0.4}{
\includegraphics[width=16.0cm,clip]{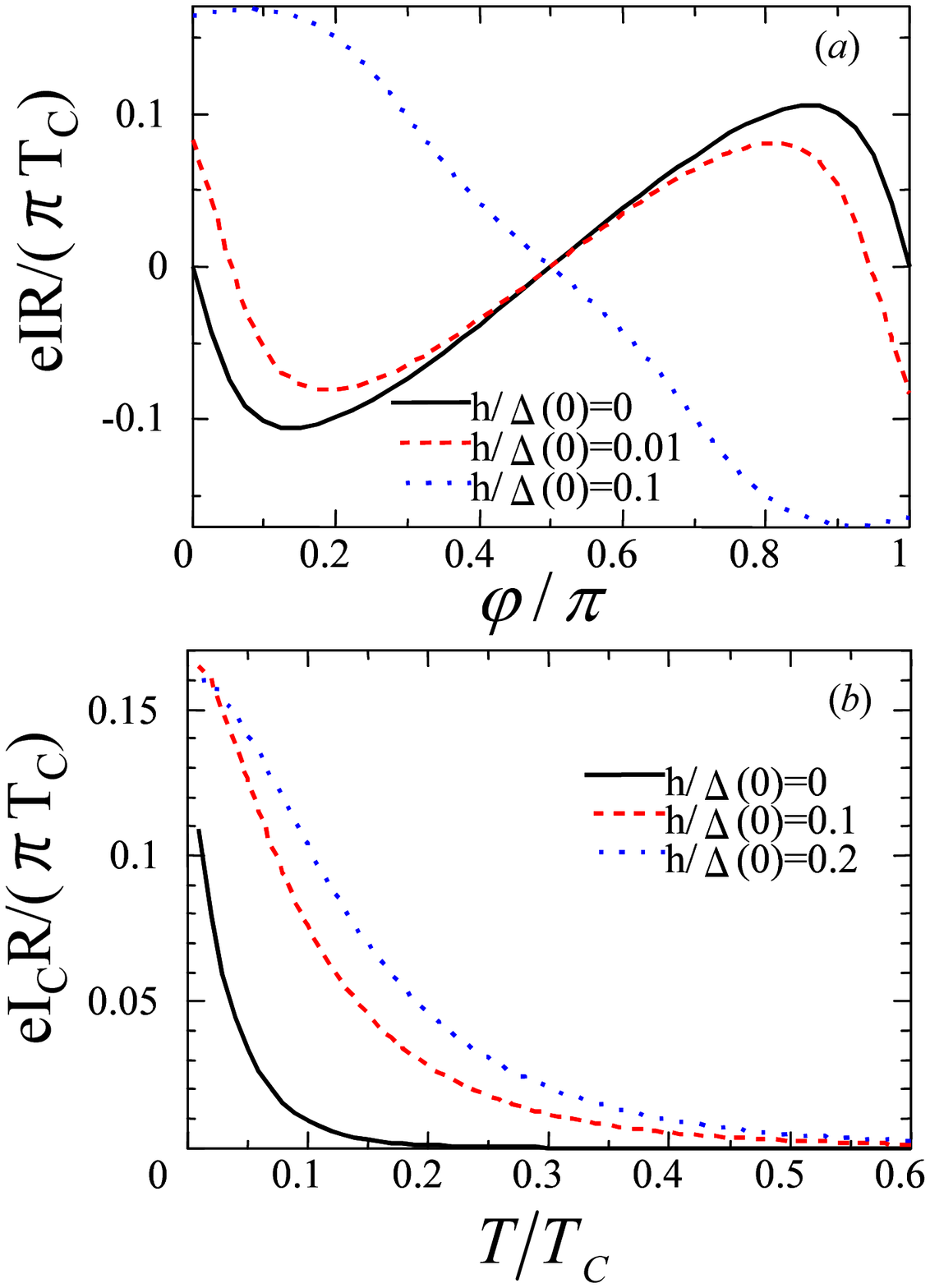}}
\end{center}
\caption{ (color online) (a) Current-phase relation at $T/T_{C}=0.01$ and
(b) temperature dependence of the critical current in S/DF/P junctions for $%
Z=10$, $R_d/R_b=1, E_{Th}/\Delta(0)=0.1$ and $\protect\beta=0$. }
\label{f14}
\end{figure}

Finally we study S/DF/P junctions for $Z=10$, $R_{d}/R_{b}=1$, $%
E_{Th}/\Delta (0)=0.1$ and $\beta =0$. Current-phase relation at $%
T/T_{C}=0.01$ has the form of  $-\sin 2\varphi $ for $h=0$ due to the
difference of the parities of two superconductors\cite{TK} as shown in Fig.%
\ref{f14} (a). As $h$ increases, the shape of $IR$ transforms from $-\sin
2\varphi $ to $\cos \varphi $ since the first harmonics recover by breaking
the symmetry between up- and down- spins. Temperature dependence of the
critical current is plotted in Fig.\ref{f14} (b). The magnitude of $I_{C}R$
is enhanced by the increase of $h$ due to the recovery of the first
harmonics, in contrast to junctions between superconductors with equal
parities.

\section{Conclusions}

In this paper, we studied the Josephson effect in junctions between
unconventional superconductors with diffusive barriers. The Usadel equations
in the barrier region were solved with the generalized boundary conditions applicable to the unconventional superconductors  at the interfaces. Applying
these boundary conditions, we calculated the Josephson current in various
types of junctions: S/DN/S, D/DN/D, P/DN/P, S/DN/D, D/DF/D, P/DF/P and
S/DF/P junctions. Our main conclusions can be summarized as follows.

1. The dependences of Josephson current on the interface barrier strength $Z$
are different for S/DN/S, D/DN/D, and P/DN/P junctions. Josephson current is
suppressed by the increase of $Z$ in S/DN/S junctions while it is enhanced
by the increase of $Z$ in D/DN/D and P/DN/P junctions. In D/DN/D and P/DN/P
junctions, proximity effect is enhanced by the increase of $Z$ due to the
cancellation of the positive and negative parts of pair potential in the
angular averaging and the coexistence of MARS and proximity effect
respectively. The coexistence also induces anomalous current-phase relation
in P/DN/P junctions. When proximity effect is absent at one interface, the
second harmonics dominate in D/DN/D and P/DN/P junctions. The competition
between MARS and proximity effect causes a nonmonotonic temperature
dependence of the critical current in D/DN/D junctions. Similar dependence
can be seen in S/DN/D junctions.

2. In S/DN/S, D/DN/D and P/DN/P junctions, the critical current has an
exponential dependence on the length of the DN. The prefactor of the length
is independent of the MARS.

3. In D/DF/D, P/DF/P and S/DF/P junctions, the $\pi $ -state can be
realized. A double peak structure in temperature dependence of the critical
current occurs in D/DF/D junctions due to 0-$\pi $ transition and the
competition between MARS and proximity effect. In S/DF/P junctions, the
Josephson current can be enhanced by the exchange field, in contrast to
other types of junctions, due to the recovery of the first harmonics.

4. In D/DF/D and P/DF/P junctions, the critical current has an oscillatory
behavior as a function of the length of the DF. The period of the
oscillation becomes short for large exchange field while it is independent
of the MARS. The second harmonics show almost half periodicity compared to
the first harmonics.

%=======================================================
%\section{Results}
%=======================================================
%%%%%%%%%%%%%%%%%%%%%%%%%%%%%%%%%%%%%%%%%%%%%%%%%%%
%
%==================================================
%The authors appreciate useful and fruitful discussions with J.
%Inoue, Yu. Nazarov and H. Itoh.
T. Y. acknowledges support by JSPS Research Fellowships for Young
Scientists. This work was supported by NAREGI Nanoscience Project, the
Ministry of Education, Culture, Sports, Science and Technology, Japan, the
Core Research for Evolutional Science and Technology (CREST) of the Japan
Science and Technology Corporation (JST), the Grant-in-Aid for Scientific
Research on Priority Area "Novel Quantum Phenomena Specific to Anisotropic
Superconductivity" (Grant No. 17071007) from the Ministry of Education,
Culture, Sports, Science and Technology of Japan, the Grant-in-Aid for
Scientific Research on B (Grant No. 17340106) from the Ministry of
Education, Culture, Sports, Science and Technology of Japan, and for the
21st Century COE "Frontiers of Computational Science" . The computational
aspect of this work has been performed at the Research Center for
Computational Science, Okazaki National Research Institutes and the
facilities of the Supercomputer Center, Institute for Solid State Physics,
University of Tokyo and the Computer Center.

%NAREGI Nanoscience Project, the Ministry of Education, Culture,
%Sports, Science and Technology, Japan, the Core Research for Evolutional
%Science and Technology (CREST) of the Japan Science and Technology
%Corporation (JST) and a Grant-in-Aid for the 21st Century COE "Frontiers of
%Computational Science" . The computational aspect of this work has been
%performed at the Research Center for Computational Science, Okazaki National
%Research Institutes and the facilities of the Supercomputer Center,
%Institute for Solid State Physics, University of Tokyo and the Computer Center
%======Reference===================================
%

%===============================================================

\end{document}